\documentclass[11pt]{article}

\usepackage{sectsty}
\sectionfont{\sc}
\renewcommand\thesection{\Roman{section}.}
\renewcommand\thesubsection{\Alph{subsection}.}

\usepackage{graphicx}
\usepackage{citesort}
\usepackage{amssymb}
\usepackage{amsfonts}
\usepackage{amsmath}
\usepackage{epsfig}
\usepackage{fancybox}
\usepackage{textcomp}
\usepackage{multirow}
\usepackage{appendix}
\usepackage{setspace}
\usepackage{subfigure}

\usepackage[top=1.3cm,bottom=1.3cm,left=1.5cm,right=1.5cm]{geometry}
\renewcommand{\baselinestretch}{1.75}
    \def\Complex{{\rm\rule[.23ex]{.03em}{1.1ex}\kern-.3em{C}}}

    \newcommand{\be}{\begin{equation}} \newcommand{\ee}{\end{equation}}
    \newcommand{\bea}{\begin{eqnarray}} \newcommand{\eea}{\end{eqnarray}}
    \newcommand{\benum}{\begin{enumerate}} \newcommand{\eenum}{\end{enumerate}}
    \def\proof{\noindent\hspace{2em}{\itshape Proof: }}
    
    \def\endproof{\hspace*{\fill}~$\square$\par\endtrivlist\unskip}

  \newtheorem{theorem}{Theorem}

  \newtheorem{lemma}{Lemma}

\newtheorem{Corollary}{Corollary}

\begin{document}

\title{\Huge\bf Fluid Antenna Systems\thanks{The work is supported by EPSRC under grant EP/M016005/1.}}

\setcounter{footnote}{2}

\author{Kai-Kit Wong\thanks{Department of Electronic and Electrical Engineering, University College London, London, United Kingdom.}, Arman Shojaeifard\thanks{BT Labs, Ipswich, United Kingdom.}, Kin-Fai Tong$\stackrel{\S}{,}$ and Yangyang Zhang\thanks{Kuang-Chi Institute of Advanced Technology, Shenzhen, China.}}

\maketitle
\thispagestyle{empty}

\begin{abstract}
Over the past decades, multiple antenna technologies have appeared in many different forms, most notably as multiple-input multiple-output (MIMO), to transform wireless communications for extraordinary diversity and multiplexing gains. The variety of technologies has been based on placing a number of antennas at fixed locations which dictates the fundamental limit on the achievable performance. By contrast, this paper envisages the scenario where the physical position of an antenna can be switched freely to one of the $N$ positions over a fixed-length line space to pick up the strongest signal in the manner of traditional selection combining. We refer to this system as a {\em fluid antenna system} (FAS) for tremendous flexibility in its possible shape and position. The aim of this paper is to study the achievable performance of a single-antenna FAS system with a fixed length and $N$ in arbitrarily correlated Rayleigh fading channels. Our contributions include exact and approximate closed-form expressions for the outage probability of FAS. We also derive an upper bound for the outage probability, from which it is shown that a single-antenna FAS given any arbitrarily small space can outperform an $L$-antenna maximum ratio combining (MRC) system if $N$ is large enough. Our analysis also reveals the minimum required size of the FAS, and how large $N$ is considered enough for the FAS to surpass MRC.

\begin{center}
{\bf Index Terms}
\end{center}
Fluid antennas, MIMO, Multiple antennas, Selection combining, Outage.
\end{abstract}

\thanks{\singlespacing}
\setcounter{page}{0}

\newpage
\section{Introduction}
Wireless communications technologies have come a long way and there is never an absence of innovative technologies when a new generation is introduced to succeed the current generation \cite{Weldon-2014}. Nonetheless, few would disagree that the most celebrated wireless technology over the past thirty years has to be multiple-input multiple-output (MIMO). The phenomenon that MIMO can create bandwidth out of space, independent of time and frequency resources, has shocked numerous researchers at the time and ever since revolutionized wireless communications technologies via many forms, e.g., \cite{Paulraj-1994,Foschini-1998,Varokh-1998,Alamouti-1998,Tse-2003,Goldsmith-2003,Spencer-2004,Marzetta-2013,Marzetta-2014}. 

The earliest version of MIMO emerged in the patent by Paulraj and Kailath in 1994 \cite{Paulraj-1994} but it was Foschini and Gans' work in \cite{Foschini-1998} that marked the beginning of MIMO research for years to come, as their paper provided a theoretical argument, for the first time, that MIMO is capable of scaling up capacity with the number of antennas at both ends. Around the same time, space-time codes by Tarokh {\em et al.}~\cite{Varokh-1998} and Alamouti \cite{Alamouti-1998} further laid a solid foundation of MIMO. The intuition that multipath can boost capacity performance was not easy to swallow at the time, and the capacity benefit of MIMO was only formally characterized by introducing the notion of diversity and multiplexing gains by Zheng and Tse in 2003 \cite{Tse-2003}. Since then, MIMO has continued to influence the development of wireless communications technology, taking the form of multiuser MIMO \cite{Goldsmith-2003,Spencer-2004} and massive MIMO being its latest version \cite{Marzetta-2013,Marzetta-2014}.

The rationale behind MIMO is to exploit diversity over multiple signals at the antennas undergoing independent fading, and the different fading signatures enabled by MIMO allow data to be multiplexed and more data to be conveyed over wireless medium. Spatial correlation between antennas is understood to be undesirable for spatial diversity of MIMO. At the mobile devices such as tablets and handsets, etc., the rule of thumb has been to only deploy antennas with at least half-wavelength apart to have the full diversity benefit. This rule may be justified by recognizing the fact that the correlation between two antennas with separation $d$ is expected to follow the Jake's model \cite{Stuber-2002}, so that
\begin{equation}
J_0\left(\frac{2\pi d}{\lambda}\right)=0\Rightarrow d=\left(\frac{2.4}{2\pi}\right)\lambda\approx 0.38\lambda\approx \frac{\lambda}{2},
\end{equation}
where $J_0(\cdot)$ is the zero-order Bessel function of the first kind, and $\lambda$ is the wavelength of propagation. Another reason why it is not advisable to fit antennas with less than half-wavelength apart is that when antennas are too close, they start interacting in the near field, giving rise to coupling. In light of this, the use of MIMO at mobile devices is greatly limited by the available space, not to mention the hardware cost of an RF chain for each antenna that is added to the system.

\begin{figure}[]
\begin{center}
\subfigure[A typical received signal across a space of $2\lambda$]{\includegraphics[width=15cm]{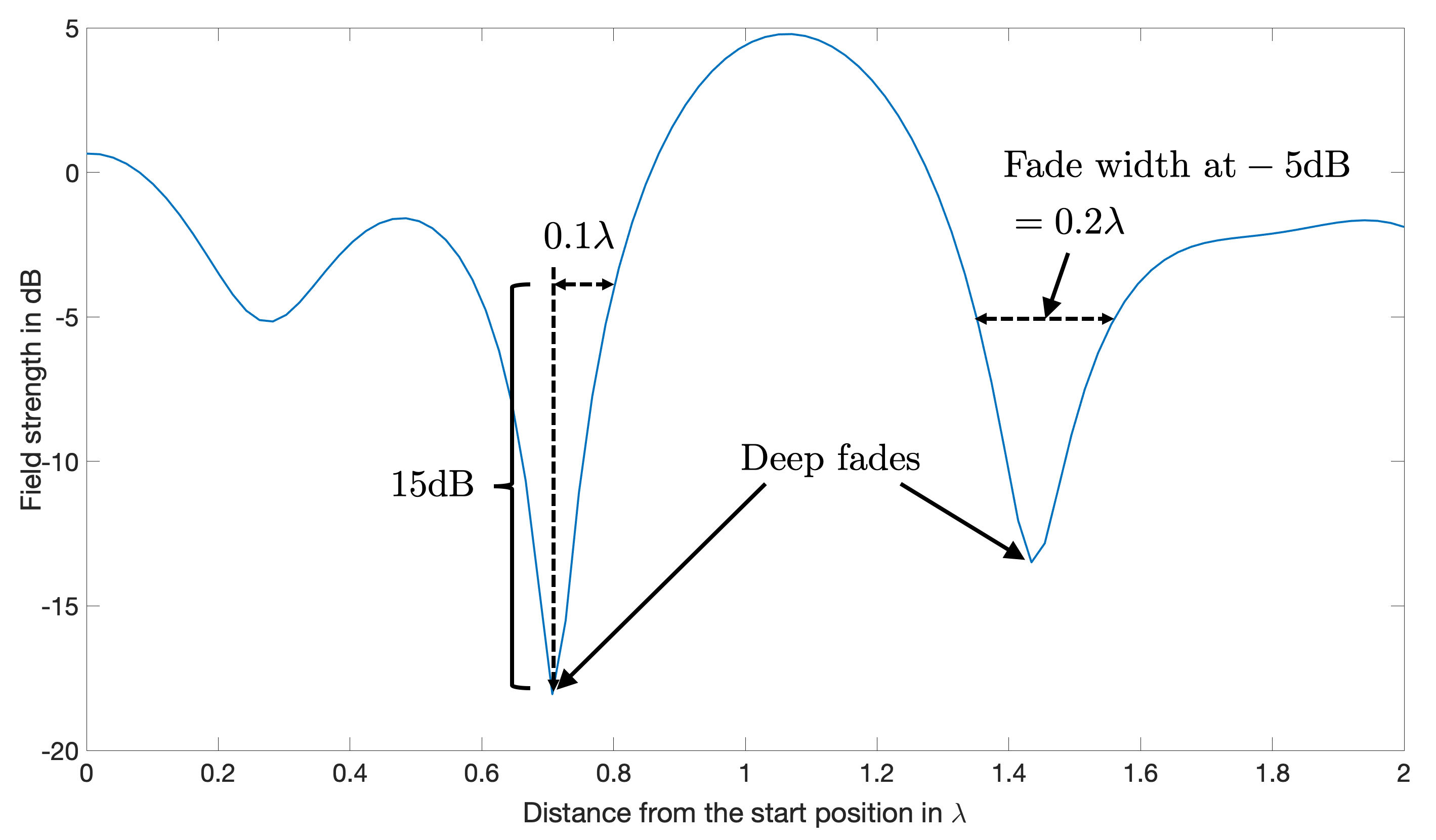}\label{fig:afd-x}}
\subfigure[Comparison for a $100$-port FAS with $2\lambda$ space and MRC at 5GHz at $v=30$km/h]{\includegraphics[width=18cm]{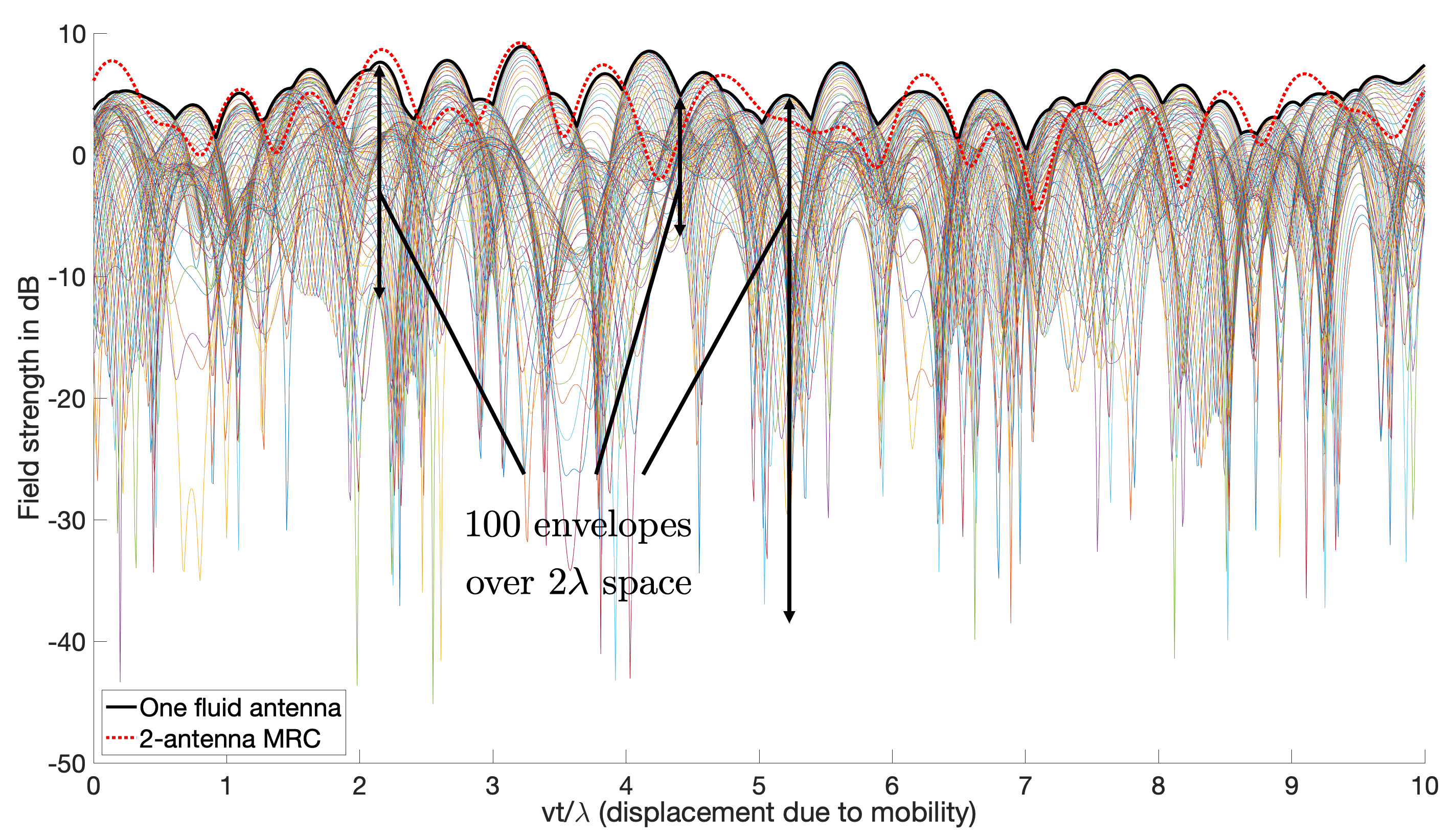}\label{fig:afd-t}}
\caption{Examples for the fading envelopes.}\label{fig:afd}
\end{center}
\end{figure}

Space at mobile devices is precious. Decades of effort have been devoted to designing antennas with high gains, large bandwidth, and agility as well as a small form factor. Conventional antennas are mainly based on metallic elements with high conductivity, strong mechanical stability, and demonstrated successes in various reconfigurability \cite{Randy-2013}. Although antennas do get much smaller over the years, advances in antenna technologies contribute only little to the massive deployment of antennas at the mobile side, as long as the $\frac{\lambda}{2}$-rule remains. This paper argues that significant performance enhancement is possible even with antennas located very close to each other in a tiny space. This can be comprehended by observing the fading behaviour of the envelope of a typical received signal, as shown in Figure \ref{fig:afd-x}. The deeper the fade, the shorter the displacement required to get out of the fade. As small as $\frac{\lambda}{10}$ in space could make the difference between at the bottom of a deep fade and a plateau of great reception.

All things considered, although it is not cost effective to deal with the complicated issue of mutual coupling and fit many antennas within a small space for diversity, antennas at highly correlated spaces do potentially benefit in the mitigation of fading. In this paper, we hypothesize a system where a {\em single} antenna can switch locations instantly in a small linear space, and refer to this system as a {\em fluid antenna system (FAS)} for its dynamic nature in possible shape and position. This paradigm is motivated by the recent trend of using liquid metals such as Galinstan, Eutectic Gallium-Indium (EGaIn) and etc., or more accessible ionized solutions such as sodium chloride (NaCl) and potassium chloride (KCl) for antennas, e.g., \cite{Hayes-2012,Ohta-2013,Saghati-2014,Dey-2016,Tong-2017,Tong-2018,Singh-2019,Tong-2019}; see \cite{Soh-2020} for a review of liquid metal antennas. While this is very much a growing research area, there is no doubt that software-controlled, mechanically flexible antennas are on their way.

Inspired by this, this paper considers that the FAS has $N$ fixed locations (known as `ports'),\footnote{It is worth pointing out that a port in 3GPP normally means an antenna TXRU (RF chain) with a reference signal but for the FAS, all ports share one single TXRU (RF chain).} evenly distributed along a linear dimension of a given length, where the mechanically flexible antenna can always be switched to the port with the strongest signal.\footnote{Note that while we are inspired by the research and use the term `fluid antenna' in this paper, the results of our work are also applicable to other flexible antenna structures such as software-controlled pixel antennas \cite{Murch-2014}.} Of particular relevance is the situation in which the space making up the FAS is small but $N$ can be very large, corresponding to the case that diversity is obtained from a large number of spatially correlated ports. It is worth noting that mutual coupling is not an issue in our model because only one antenna exists. The aim of this paper is to investigate the theoretical performance of such FAS and quantify its benefit in terms of outage probability performance. Statistical analysis will be our main tool to make possible derivation of the outage performance for the $N$-port FAS, but our emphasis is to incorporate the roles of various physical parameters such as the size of the space, and the number of ports, and examine how they impact the performance.

In summary, this paper has made the following contributions:
\begin{itemize}
\item We derive the joint probability density function (pdf) and cumulative density function (cdf) of the envelopes of the signals at all the antenna ports for the FAS, under spatially-correlated Rayleigh fading channels. The exact outage probability for the FAS is derived in a single integral form.
\item In addition, we provide an approximate outage probability expression for the FAS in closed form, which is proven tight for harsh situations with stringent targets and strong fading correlations.
\item Also, we propose an upper bound for the outage probability which permits us to conduct several important comparative studies. First, we show that as $N\to\infty$, the FAS with any given dimension, however small, can achieve any arbitrarily small outage probability. Secondly, we provide an analytical quantification to work out how large $N$ is considered enough for the single-antenna FAS to outperform an $L$-antenna maximum ration combining (MRC) system. Furthermore, for large but finite $N$, we obtain the minimum required size of the FAS for surpassing MRC.
\end{itemize}

The remainder of the paper is organized as follows. In Section II, we present the system model of an $N$-port FAS. Then the main results of this paper are presented in Section III which includes the joint pdf and cdf of the signal envelopes of the antenna ports, the exact and approximate outage probability expressions, and an outage probability upper bound. Several analysis characterizing the impact of the system parameters is also provided in Section III. Section IV presents the the numerical results corroborating the analytical results we derive. Finally, we conclude the paper in Section V.


\section{System Model}\label{sec:model}
\subsection{Physical Model}\label{ssec:pm}
\begin{figure}[]
\begin{center}
\includegraphics[width=13cm]{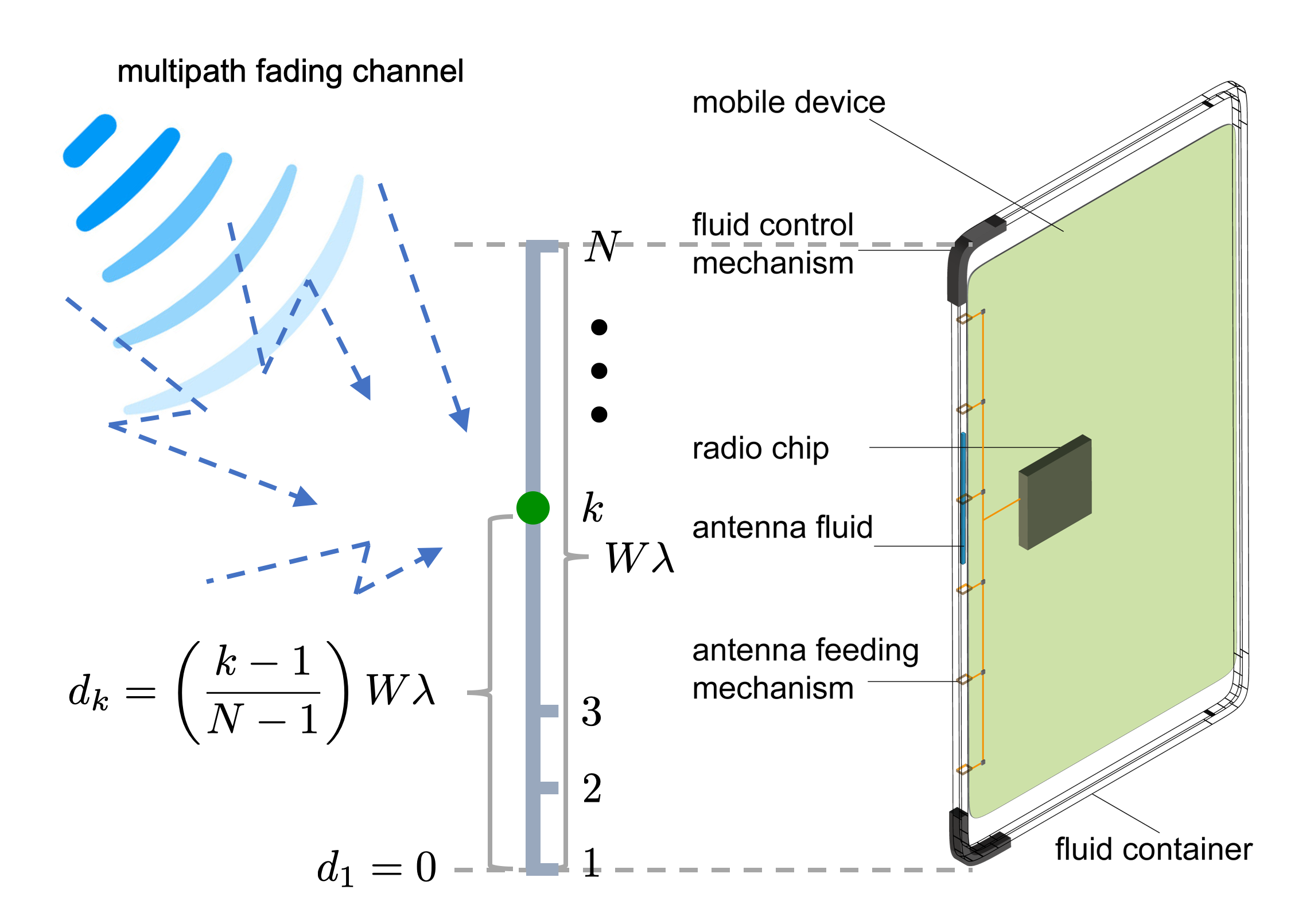}
\caption{A possible architecture for a FAS.}\label{fig:fluid_antenna}
\end{center}
\end{figure}

In this paper, we envisage a single-antenna wireless device for reception where the antenna's location can be switched to one of the $N$ preset locations evenly distributed along a linear dimension of length, $W\lambda$, at the chassis of the wireless device, where $\lambda$ denotes the wavelength of radiation. Figure \ref{fig:fluid_antenna} depicts one such architecture that could realize the mobilization of a physical antenna made of liquid metal or ionized solutions \cite{Hayes-2012,Ohta-2013,Saghati-2014,Dey-2016,Tong-2017,Tong-2018,Singh-2019,Tong-2019,Soh-2020}. We consider an abstraction of the FAS concept where an antenna at a given location (referred to as a `port') is treated as an ideal point antenna. As shown in Figure \ref{fig:fluid_antenna}, the first port is the reference location, from which the displacement of the $k$-th port is measured, denoted by
\begin{equation}\label{eqn:dn}
d_k=\left(\frac{k-1}{N-1}\right)W\lambda,~\mbox{for }k=1,2,\dots,N.
\end{equation}

The received signal at the $k$-th port is modelled as
\begin{equation}\label{eqn:yk}
y_k=g_k x+\eta_k,
\end{equation}
where the time index is omitted for conciseness, $g_k$ is the flat fading channel coefficient undergone by the $k$-th port, which is assumed to follow a circularly symmetric complex Gaussian distribution with zero mean and variance of $\sigma^2$, $\eta_k$ denotes the complex additive white Gaussian noise (AWGN) at the $k$-th port with zero mean and variance of $\sigma_\eta^2$, and $x$ denotes the transmitted data symbol. Under this model, the amplitude of the channel, $|g_k|$, is Rayleigh distributed, with the pdf
\begin{equation}\label{eqn:|g|}
p_{|g_k|}(r)=\frac{2r}{\sigma^2}e^{-\frac{r^2}{\sigma^2}},~\mbox{for }r\ge 0~\mbox{with }{\rm E}[|g_k|^2]=\sigma^2.
\end{equation}

The average received signal-to-noise ratio (SNR) at each port is given by
\begin{equation}\label{eqn:snr}
\Gamma=\sigma^2\frac{{\rm E}[|x|^2]}{\sigma^2_\eta}\equiv\sigma^2\Theta,~\mbox{where }\Theta\triangleq\frac{{\rm E}[|x|^2]}{\sigma^2_\eta}.
\end{equation}
The channels $\{g_k\}_{\forall n}$ are considered to be correlated because they can be arbitrarily close to each other. 

\subsection{Temporal Correlation Model}\label{ssec:tcm}
As the wireless device moves at speed $v$, the channel changes but Doppler effect causes it to correlate over time. With 2-D isotropic scattering and an isotropic receiver port, it is known that the autocorrelation functions of the channel satisfy \cite{Stuber-2002}
\begin{equation}\label{eqn:time-correlation}
\phi_{gg}(\tau)=\phi_{{\rm Re}\{g\}{\rm Re}\{g\}}(\tau)=\phi_{{\rm Im}\{g\}{\rm Im}\{g\}}(\tau)=\frac{\sigma^2}{2}J_0(2\pi f_m\tau)
\end{equation}
where $\tau$ represents the time difference, and $f_m=\frac{v}{\lambda}$ is the maximum Doppler frequency.

\subsection{Spatial Correlation Model}\label{ssec:scm}
For antenna ports in FAS, the spatial separation between ports, $\Delta d$, analogous to $v\tau$ in the temporal correlation case as a result of moving in Section II-B, leads to phase difference of the arriving paths, and hence results in correlation of the channels between ports. As a result,
\begin{equation}\label{eqn:space-correlation}
\phi_{g_kg_\ell}(\Delta d_{k,\ell})=\frac{\sigma^2}{2}J_0\left(2\pi\frac{\Delta d_{k,\ell}}{\lambda}\right)=\frac{\sigma^2}{2}J_0\left(\frac{2\pi(k-\ell)}{N-1}W\right).
\end{equation}

\subsection{Statistical Model for Single-Antenna $N$-Port FAS}\label{ssec:fas}
For ease of exposition, we parameterize the channels at the $N$ antenna ports in a FAS by 
\begin{equation}\label{eqn:g-model}
\left\{\begin{aligned}
g_1&=\sigma x_0 +j\sigma y_0\\
g_2&=\sigma\left(\sqrt{1-\mu_2^2}x_2+\mu_2 x_0\right)+j\sigma\left(\sqrt{1-\mu_2^2}y_2+\mu_2 y_0\right)\\
& \vdots\\
g_N&=\sigma\left(\sqrt{1-\mu_N^2}x_N+\mu_N x_0\right)+j\sigma\left(\sqrt{1-\mu_N^2}y_N+\mu_N y_0\right),
\end{aligned}\right.
\end{equation}
where $x_0,x_1,\dots,x_N,y_0,y_1,\dots,y_N$ are all independent Gaussian random variables with zero mean and variance of $\frac{1}{2}$, and $\{\mu_k\}$ are the parameters that can be chosen freely to control the correlation between the channels $\{g_k\}$. Based on this model, ${\rm E}[|g_k|^2]=\sigma^2$ for all $k$ and due to (\ref{eqn:space-correlation}), we choose
\begin{equation}\label{eqn:mu-condition}
\mu_k=J_0\left(\frac{2\pi(k-1)}{N-1}W\right),~\mbox{for }k=2,\dots,N.
\end{equation}
We assume that the FAS can always switch to the maximum of $\{|g_k|\}$ instantly for the best performance\footnote{This is potentially a strong assumption as switching physical materials from one space to another causes delay although the reaction time can be much faster than normally thought; see recent advances in \cite{Neil-2019}, which is particularly true when the antenna size is getting smaller at higher frequencies. Despite this, the goal of this paper is to look at the fundamental performance limit while research efforts are being carried out to address various implementation problems. In addition, there is also the possibility that a FAS is realized by an array of mini pixels that can be digitally controlled to effectively make an antenna appear and disappear at spaces without any noticeable delay \cite{Murch-2014}.} and therefore we are interested in the random variable
\begin{equation}
|g_{\rm FAS}|=\max\{|g_1|,|g_2|,\dots,|g_N|\},
\end{equation}
in which the correlation among $\{|g_k|\}$ is specified by (\ref{eqn:mu-condition}). The cdf and pdf of a random variable of this type have been reported in \cite[(5) \& (10)]{Ugweje-1997}. Nevertheless, the expressions in \cite{Ugweje-1997} are in the form of multiple integrals, which prohibit any insightful analysis. Section III will revisit this and derive exact expressions for the cdf and pdf of $|g_{\rm FAS}|$ and the outage probability in the forms that we can gain useful insight and link the outage performance with the physical parameters of FAS.


\section{Main Results}
This section studies the statistical properties of $|g_{\rm FAS}|$ and analyzes it through the derivation of joint probability distributions of $|g_1|,|g_2|,\dots,|g_N|$, and outage probability expressions.

\subsection{Probability Distributions and Outage Probability}
\begin{theorem}\label{th:jpdf}
The joint pdf of $|g_1|,|g_2|,\dots,|g_N|$ is given by
\begin{equation}\label{eqn:jpdf}
p_{|g_1|,|g_2|,\dots,|g_N|}(r_1,r_2,\dots,r_N)=\prod_{k=1\atop (\mu_1\triangleq 0)}^N\frac{2r_k}{\sigma^2(1-\mu_k^2)}e^{-\frac{r_k^2+\mu_k^2r_1^2}{\sigma^2(1-\mu_k^2)}}I_0\left(\frac{2\mu_kr_1r_k}{\sigma^2(1-\mu_k^2)}\right),~\mbox{for }r_1,\dots,r_N\ge 0,
\end{equation}
where $I_0(\cdot)$ is the zero-order modified Bessel function of the first kind.
\end{theorem}

\proof See Appendix A.
\endproof

\begin{Corollary} 
For $N=2$, the joint pdf of $|g_1|,|g_2|$ becomes
\begin{equation}
p_{|g_1|,|g_2|}(r_1,r_2)=\frac{4r_1r_2}{\sigma^4(1-\mu_2^2)}e^{-\frac{r_1^2}{\sigma^2}}e^{-\frac{r_2^2+\mu_2^2r_1^2}{\sigma^2(1-\mu_2^2)}}I_0\left(\frac{2\mu_2r_1r_2}{\sigma^2(1-\mu_2^2)}\right),~\mbox{for }r_1,r_2\ge 0.
\end{equation}
\end{Corollary}

\proof This result comes directly from Theorem \ref{th:jpdf} if $N=2$ and agrees with the result in \cite{Slim-2003}.
\endproof

\begin{theorem}\label{th:jcdf}
The joint cdf of $|g_1|,|g_2|,\dots,|g_N|$ is given by
\begin{align}
F_{|g_1|,|g_2|,\dots,|g_N|}(r_1,r_2,\dots,r_N)&={\rm Prob}(|g_1|<r_1,|g_2|<r_2,\dots,|g_N|<r_N)\notag\\
&=\int_0^{\frac{r_1^2}{\sigma^2}}e^{-t}\prod_{k=2}^N\left[1-Q_1\left(\sqrt{\frac{2\mu_k^2}{1-\mu_k^2}}\sqrt{t},\sqrt{\frac{2}{\sigma^2(1-\mu_k^2)}}r_k\right)\right]dt,\label{eqn:jcdf}
\end{align}
where $Q_1(a,b)$ denotes the first-order Marcum $Q$-function \cite{Simon-2002}.
\end{theorem}

\proof See Appendix B.
\endproof

\begin{Corollary} 
For $N=2$, the joint cdf of $|g_1|,|g_2|$ can be found as
\begin{multline}\label{eqn:jcdf2}
F_{|g_1|,|g_2|}(r_1,r_2)=1-e^{-\frac{r_1^2}{\sigma^2}}-e^{-\frac{r_2^2}{\sigma^2}}Q_1\left(\sqrt{\frac{2}{\sigma^2(1-\mu_2^2)}}r_1,\sqrt{\frac{2\mu_2^2}{\sigma^2(1-\mu_2^2)}}r_2\right)\\
-e^{-\frac{r_1^2}{\sigma^2}}Q_1\left(\sqrt{\frac{2\mu_2^2}{\sigma^2(1-\mu_2^2)}}r_1,\sqrt{\frac{2}{\sigma^2(1-\mu_2^2)}}r_2\right).
\end{multline}
\end{Corollary}

\proof See Appendix C.
\endproof

With the joint cdf obtained in Theorem \ref{th:jpdf}, we present the following theorem to compute the outage probability of the FAS by defining the outage event  as $\{|g_{\rm FAS}|^2\Theta<\gamma_{\rm th}\}=\left\{|g_{\rm FAS}|<\sqrt{\frac{\gamma_{\rm th}}{\Theta}}\right\}$.

\begin{theorem}\label{th:op}
The outage probability of the FAS is given by
\begin{equation}\label{eqn:op}
p_{\rm out}(\gamma_{\rm th})=\int_0^{\frac{\gamma_{\rm th}}{\Gamma}}e^{-t}\prod_{k=2}^N\left[1-Q_1\left(\sqrt{\frac{2\mu_k^2}{1-\mu_k^2}}\sqrt{t},\sqrt{\frac{2}{1-\mu_k^2}}\sqrt{\frac{\gamma_{\rm th}}{\Gamma}}\right)\right]dt.
\end{equation}
\end{theorem}

\proof The outage probability expression can be found by substituting $r_1=r_2=\cdots=r_N=\sqrt{\frac{\gamma_{\rm th}}{\Theta}}$ into the joint cdf (\ref{eqn:jcdf}) and using the average SNR definition in (\ref{eqn:snr}) which completes the proof.
\endproof

\begin{Corollary} 
For $N=2$, the outage probability is expressed as
\begin{equation}\label{eqn:op2}
p_{\rm out}^{N=2}(\gamma_{\rm th})=1-e^{-\frac{\gamma_{\rm th}}{\Gamma}}-e^{-\frac{\gamma_{\rm th}}{\Gamma}}\Delta Q_1\left(\sqrt{\frac{2}{1-\mu_2^2}}\sqrt{\frac{\gamma_{\rm th}}{\Gamma}},\sqrt{\frac{2\mu_2^2}{1-\mu_2^2}}\sqrt{\frac{\gamma_{\rm th}}{\Gamma}}\right),
\end{equation}
where $\Delta Q_1(\alpha,\beta)\triangleq Q_1(\alpha,\beta)-Q_1(\beta,\alpha)$. Note that the last term quantifies the exact benefit of the second antenna port for reduction of the outage probability which depends on the autocorrelation $\mu_2$.
\end{Corollary}

\proof Setting $r_1=r_2=\sqrt{\frac{\gamma_{\rm th}}{\Theta}}$ in (\ref{eqn:jcdf2}) and using the definition of $\Delta Q_1(\alpha,\beta)$ and that of the average SNR in (\ref{eqn:snr}) give the desired result.
\endproof

As a sanity check, we consider two special cases below. The first case concerns $p_{\rm out}^{N=2}(\gamma_{\rm th})$ when $\mu_2=0$ (i.e., the two ports are independent). In this case, $p_{\rm out}^{N=2}(\gamma_{\rm th})$ can be simplified as
\begin{equation}
p_{\rm out}^{N=2}(\gamma_{\rm th})=1-e^{-\frac{\gamma_{\rm th}}{\Gamma}}-e^{-\frac{\gamma_{\rm th}}{\Gamma}}\left[Q_1\left(\sqrt{2}\sqrt{\frac{\gamma_{\rm th}}{\Gamma}},0\right)-Q_1\left(0,\sqrt{2}\sqrt{\frac{\gamma_{\rm th}}{\Gamma}}\right)\right].
\end{equation}
Using the facts that $Q_1(\alpha,0)=1$ and $Q_1(0,\beta)=e^{-\frac{\beta^2}{2}}$, we get
\begin{equation}
p_{\rm out}^{N=2}(\gamma_{\rm th})=1-e^{-\frac{\gamma_{\rm th}}{\Gamma}}-e^{-\frac{\gamma_{\rm th}}{\Gamma}}\left(1-e^{-\frac{\gamma_{\rm th}}{\Gamma}}\right)=\left(1-e^{-\frac{\gamma_{\rm th}}{\Gamma}}\right)^2,
\end{equation}
showing that a diversity order of $2$ is obtained, as expected.

The second case considers $\mu_2=1$ which corresponds to the situation where the two ports are identical. Before we proceed to analyze $p_{\rm out}^{N=2}(\gamma_{\rm th})$, the following lemma is useful.

\begin{lemma}\label{lemma:lim1}
\begin{equation}
\lim_{\alpha\to\infty}Q_1(\alpha,\alpha)=\frac{1}{2}.
\end{equation}
\end{lemma}

\proof To prove the result, we first consider that for large $y>0$, $I_0(y)=\frac{\xi e^y}{\sqrt{2\pi y}}$ for some constant $\xi$. Thus, we have $e^{-y}I_0(y)=\frac{\xi}{\sqrt{2\pi y}}\to 0$ as $y\to\infty$. Then it is known that \cite{Simon-2002}
\begin{equation}
Q_1(\alpha,\alpha)=\frac{1}{2}\left(1+e^{-\alpha^2}I_0(\alpha^2)\right),
\end{equation}
which goes to $\frac{1}{2}$ as $\alpha\to\infty$ because $\lim_{y\to\infty}e^{-y}I_0(y)=0$.
\endproof

Using the result of Lemma \ref{lemma:lim1}, we are then able to obtain, for the case of $\mu_2=1$, that
\begin{equation}
p_{\rm out}^{N=2}(\gamma_{\rm th})=1-e^{-\frac{\gamma_{\rm th}}{\Gamma}}-e^{-\frac{\gamma_{\rm th}}{\Gamma}}\left(\frac{1}{2}-\frac{1}{2}\right)=1-e^{-\frac{\gamma_{\rm th}}{\Gamma}}.
\end{equation}
As expected, this case is reduced to the single-antenna case with no diversity.

The $N=2$ case, however, does not capture the concept of FAS well. The case with large $N$ is more relevant to unleash the true potential of FAS.\footnote{It is worth pointing out that in traditional multiple antenna systems, it is a common practice that more than one antennas are equipped only if they are spaced at least $\frac{\lambda}{2}$ apart for the full benefit of diversity. By contrast, under FAS, regardless of how small the space is, less than $\frac{\lambda}{2} (\mbox{i.e., } W<0.5)$ or not, $N$ can still be very large as the fluid antenna can freely switch its location within the space of $W\lambda$. Our analysis in Section III-B will reveal that FAS can deliver remarkable performance even with a small $W$. Also, note that only one RF chain is required for FAS regardless of the value of $N$.} Although (\ref{eqn:op}) in Theorem \ref{th:op} permits us to evaluate the exact outage probability via a single integral with a finite interval, the expression is not very useful in understanding how the performance of FAS scales with $N$ and other system parameters. To tackle this, we first obtain an approximation of the outage probability in closed form.

\begin{theorem}\label{th:aop}
The outage probability of the FAS can be approximated by
\begin{equation}\label{eqn:aop}
p_{\rm out}(\gamma_{\rm th})\approx 1-e^{-\frac{\gamma_{\rm th}}{\Gamma}}-e^{-\frac{\gamma_{\rm th}}{\Gamma}}\sum_{k=2}^N\Delta Q_1\left(\sqrt{\frac{2}{1-\mu_k^2}}\sqrt{\frac{\gamma_{\rm th}}{\Gamma}},\sqrt{\frac{2\mu_k^2}{1-\mu_k^2}}\sqrt{\frac{\gamma_{\rm th}}{\Gamma}}\right).
\end{equation}
\end{theorem}

\proof See Appendix D.
\endproof

The expression (\ref{eqn:aop}) is in closed form which is much easier to compute than (\ref{eqn:op}) but more crucially, it enables us to quantify the benefit of each additional antenna port for reduction of the outage probability. In particular, it can be deduced from (\ref{eqn:aop}) that each antenna port with autocorrelation parameter $\mu_k$ contributes to a reduction of $e^{-\frac{\gamma_{\rm th}}{\Gamma}}\Delta Q_1\left(\sqrt{\frac{2}{1-\mu_k^2}}\sqrt{\frac{\gamma_{\rm th}}{\Gamma}},\sqrt{\frac{2\mu_k^2}{1-\mu_k^2}}\sqrt{\frac{\gamma_{\rm th}}{\Gamma}}\right)$ in outage probability. The tightness of the approximation is addressed by the following proposition.

\begin{theorem}\label{th:aop-tight}
The outage probability approximation (\ref{eqn:aop}) is tight if $\{\mu_k\}$ and/or $\frac{\gamma_{\rm th}}{\Gamma}$ are large.
\end{theorem}

\proof See Appendix E.
\endproof

Theorem \ref{th:aop-tight} suggests that for the cases where the linear space of $W\lambda$ is small (hence, $\{\mu_k\}$ are large with strong spatial correlation), and the SNR threshold relative to the average SNR is stringent, the approximation be tight and therefore (\ref{eqn:aop}) will be useful for characterizing the performance of the FAS. However, under such very harsh condition, the outage probability is deemed to be very high, and may be too high to have a meaningful system. To this end, we derive an upper bound of the outage probability which works for typical circumstances. Before that, we first present the following theorem.

\begin{theorem}\label{th:op-delta}
The exact reduction in outage probability (\ref{eqn:op}) due to the $N$-th port for an $(N-1)$-port FAS can be determined by
\begin{multline}\label{eqn:op-delta}
\Delta p_{\rm out}(\gamma_{\rm th})=\\
\int_0^{\frac{\gamma_{\rm th}}{\Gamma}}e^{-t}Q_1\left(\sqrt{\frac{2\mu_N^2}{1-\mu_N^2}}\sqrt{t},\sqrt{\frac{2}{1-\mu_N^2}}\sqrt{\frac{\gamma_{\rm th}}{\Gamma}}\right)\prod_{k=2}^{N-1}\left[1-Q_1\left(\sqrt{\frac{2\mu_k^2}{1-\mu_k^2}}\sqrt{t},\sqrt{\frac{2}{1-\mu_k^2}}\sqrt{\frac{\gamma_{\rm th}}{\Gamma}}\right)\right]dt.
\end{multline}
\end{theorem}

\proof The result is obtained by expanding the $N$-th factor in the product of (\ref{eqn:op}).
\endproof

Noting that $Q_1(\alpha,\beta)\to 0$ as $\beta\to\infty$ if $\alpha<\beta$, we can see from (\ref{eqn:op-delta}) that $Q_1\left(\sqrt{\frac{2\mu_N^2}{1-\mu_N^2}}\sqrt{t},\sqrt{\frac{2}{1-\mu_N^2}}\sqrt{\frac{\gamma_{\rm th}}{\Gamma}}\right)$ inside the integral is zero if $\mu_N=1$ and hence $\Delta p_{\rm out}(\gamma_{\rm th})=0$. In other words, there will be no reduction in outage probability if the additional port is identical to the first port, which agrees with the intuition.

\begin{theorem}\label{th:op-ub}
An upper bound of the outage probability (\ref{eqn:op}) is given by\footnote{The upper bound is valid only if the factors inside the product operator are all positive. If for some $|\mu_{\hat{k}}|$ that is close to $0$, the factor could become negative. To mitigate this, we can replace $1-\frac{\varrho}{\sqrt{|\mu_{\hat{k}}|}}e^{-\frac{\kappa}{1-\mu_{\hat{k}}^2}\left(\frac{\gamma_{\rm th}}{\Gamma}\right)}$ by $1-\varrho e^{-\frac{\kappa}{1-\mu_{\hat{k}}^2}\left(\frac{\gamma_{\rm th}}{\Gamma}\right)}$, and the modified expression should still be an upper bound. On the other hand, the upper bound can be tighter if $p_{\rm out}^{\rm UB}(\gamma_{\rm th})=p_{\rm out}^{2}(\gamma_{\rm th})\prod_{k=3}^N(\cdot)$, where $p_{\rm out}^{2}(\gamma_{\rm th})$ can be obtained by (\ref{eqn:op2}). In the simulations, however, (\ref{eqn:op-ub}) will be used to produce the numerical results.}
\begin{equation}\label{eqn:op-ub}
p_{\rm out}(\gamma_{\rm th})<p_{\rm out}^{\rm UB}(\gamma_{\rm th})
=\left(1-e^{-\frac{\gamma_{\rm th}}{\Gamma}}\right)\prod_{k=2}^N\left(1-\frac{\varrho}{\sqrt{|\mu_k|}}e^{-\frac{\kappa}{1-\mu_k^2}\left(\frac{\gamma_{\rm th}}{\Gamma}\right)}\right),
\end{equation}
where $\kappa>1$ and $0<\varrho<0.5$ are some constants given in Appendix F.
\end{theorem}

\proof See Appendix F.
\endproof

\begin{Corollary}
Based on the upper bound, the $k$-th antenna port, if highly correlated, of the FAS contributes to a diversity gain of $\frac{\varrho}{\sqrt{|\mu_k|}}$ but it also comes with an SNR scaling penalty of $\frac{1-\mu_k^2}{\kappa}$.
\end{Corollary}

\proof From (\ref{eqn:op-ub}), it can be seen that the $k$-th antenna port contributes to a reduction in the outage probability bound by a scaling of $1-\frac{\varrho}{\sqrt{|\mu_k|}}e^{-\frac{\kappa}{1-\mu_k^2}\left(\frac{\gamma_{\rm th}}{\Gamma}\right)}\approx\left(1-e^{-\frac{\kappa}{1-\mu_k^2}\left(\frac{\gamma_{\rm th}}{\Gamma}\right)}\right)^\frac{\varrho}{\sqrt{|\mu_k|}}$ for large $|\mu_k|$. Also, comparing this expression with the scaling factor contributed by an antenna with independent fading, i.e., $\left(1-e^{-\frac{\gamma_{\rm th}}{\Gamma}}\right)^1$, the penalty on the average SNR, $\Gamma$, can be observed directly. 
\endproof

\subsection{Physical Insights}
In this subsection, our aim is to gain some physical insight that can help understand the operation of FAS in relation to its physical parameters such as its dimension and the number of ports. As a performance benchmark, we consider an $L$-antenna MRC receiver with independent fading. Our interest is to figure out under what conditions the single-antenna $N$-spatially-correlated-port FAS with a fixed dimension outperforms MRC. To start with, we establish the performance of MRC in the following lemma.

\begin{lemma}\label{lemma:mrc}
The outage probability for the $L$-antenna MRC system is given by
\begin{equation}\label{eqn:op-mrc}
p_{\rm out}^{{\rm MRC}, L}(\gamma_{\rm th})=1-e^{-\frac{\gamma_{\rm th}}{\Gamma}}\sum_{k=0}^{L-1}\frac{1}{k!}
\left(\frac{\gamma_{\rm th}}{\Gamma}\right)^k.
\end{equation}
\end{lemma}

\proof The result is directly taken from \cite[(6.25)]{Stuber-2002}.
\endproof

We now present our main results in the following theorems.

\begin{theorem}\label{th:fas-mrc}
For given SNR threshold, $\gamma_{\rm th}$, and the average SNR, $\Gamma$, the FAS with any given dimension, $W\lambda$, can achieve any arbitrarily small outage probability, if $N\to\infty$ as long as $|\mu_k|\ne 1$.
\end{theorem}

\proof As long as $|\mu_k|\ne 1$, the product in (\ref{eqn:op-ub}) is a multiplication between $N-1$ less-than-one numbers, and if $N\to\infty$, the upper bound and so the outage probability will converge to zero.
\endproof

The result in Theorem \ref{th:fas-mrc} speaks the significance of FAS as it literally confirms that a single-antenna FAS with a given dimension, however small, can achieve zero outage probability if $N$ is asymptotically large and as such, can beat an $L$-antenna MRC system with independent fading and $L$ RF chains. The next corollaries address how large $N$ needs to be for FAS in order to outperform MRC.

\begin{Corollary}\label{corollary:N-0}
For given $\{\mu_k\}$, the FAS outperforms MRC if $N$ satisfies
\begin{equation}\label{eqn:N-0}
\prod_{k=2}^N\left(1-\frac{\varrho}{\sqrt{|\mu_k|}}e^{-\frac{\kappa}{1-\mu_k^2}\left(\frac{\gamma_{\rm th}}{\Gamma}\right)}\right)<\frac{p_{\rm out}^{{\rm MRC}, L}(\gamma_{\rm th})}{1-e^{-\frac{\gamma_{\rm th}}{\Gamma}}}.
\end{equation}
\end{Corollary}

\proof The result comes directly from comparing (\ref{eqn:op-ub}) and (\ref{eqn:op-mrc}).
\endproof

\begin{Corollary}\label{corollary:N-1}
If $|\mu_2|=|\mu_3|=\cdots=|\mu_N|=\mu$, then MRC is outperformed by FAS when
\begin{equation}\label{eqn:N-1}
N>\frac{\ln\left(\frac{p_{\rm out}^{{\rm MRC}, L}(\gamma_{\rm th})}{1-e^{-\frac{\gamma_{\rm th}}{\Gamma}}}\right)}{\ln\left(1-\frac{\varrho}{\sqrt{\mu}}e^{-\frac{\kappa}{1-\mu^2}\left(\frac{\gamma_{\rm th}}{\Gamma}\right)}\right)}+1.
\end{equation}
\end{Corollary}

\proof Using (\ref{eqn:N-0}) and $|\mu_k|=\mu~\forall k$ gives the desired result.
\endproof


The following corollary obtains the required autocorrelation for the ports, $\mu$, for the FAS with a given $N$ in order to surpass MRC in the case where all the ports have the same autocorrelation parameters.

%

\begin{Corollary}\label{corollary:mu}
For the homogeneous case where $\mu_k=\mu~\forall k$ with large but finite $N$, the required autocorrelation, $\mu$, and dimension, $d$, for FAS to outperform MRC are, respectively, given by
\begin{equation}\label{eqn:mu-star}
\mu\le\mu^*\triangleq \sqrt{1-\frac{\kappa\left(\frac{\gamma_{\rm th}}{\Gamma}\right)}{\ln\left(\frac{\varrho}{1-\left(\frac{p_{\rm out}^{{\rm MRC}, L}(\gamma_{\rm th})}{1-e^{-\frac{\gamma_{\rm th}}{\Gamma}}}\right)^\frac{1}{N-1}}\right)}},~\mbox{and }d\ge d^*\triangleq\frac{J_0^{-1}(\mu^*)}{2\pi}\lambda.
\end{equation}
\end{Corollary}

\proof Changing the subject of the condition (\ref{eqn:N-1}) as $\mu$ yields the desired result for $\mu$. Then using (\ref{eqn:mu-condition}), we can find out the required dimension, $d$, of the FAS. 
\endproof

Corollary \ref{corollary:mu} reveals that the value of $N$ compensates for the spatial correlation of the antenna ports, meaning that if $N$ is larger, then $\mu$ can be closer to one, or the space required for the fluid antenna can be smaller, and the FAS still outperforms MRC in terms of outage probability. Also, theoretically in the asymptotic regime, we only need $\mu<\mu^*=1$ as $N\to\infty$, meaning that FAS with extremely correlated ports is not a problem. Corollary \ref{corollary:mu}, however, deals with only a very special case where $\mu_k=\mu~\forall k$, which may not represent any physical configuration of interest. The following theorem addresses the general case and links the dimension of the FAS, $W\lambda$, with the system parameters and $N$.


\begin{theorem}\label{th:W}
For large but finite $N$, the FAS outperforms MRC if
\begin{equation}\label{eqn:W}
W\ge\frac{1}{2\pi}J_0^{-1}\left(\sqrt{1-\frac{\kappa\left(\frac{\gamma_{\rm th}}{\Gamma}\right)}{\ln\left(\frac{\varrho}{1-\left(\frac{p_{\rm out}^{{\rm MRC}, L}(\gamma_{\rm th})}{1-e^{-\frac{\gamma_{\rm th}}{\Gamma}}}\right)^\frac{1}{\frac{N}{2}-1}}\right)}}\right).
\end{equation}
\end{theorem}

\proof See Appendix G.
\endproof

Note that for both Corollary \ref{corollary:mu} and Theorem \ref{th:W}, valid solutions may not be possible if $N$ is not large enough. This happens when the result of $\ln(\cdot)$ becomes negative, and $\mu^*>1$, or the number inside the square root operation is negative, resulting a complex $\mu^*$ if $N$ is not large enough. Also, care must be taken when performing $J_0^{-1}(\cdot)$ as $J_0(\cdot)$ is an oscillating function which goes from one to zero for distance from zero to $0.38\lambda$. Yet, $J_0(\frac{2\pi}{\lambda}\times 0.38\lambda)=0$ does not ensure that $J_0(\frac{2\pi d}{\lambda})=0$ for $d>0.38\lambda$ but (\ref{eqn:W}) relies on the fact that $|J_0(\frac{2\pi d}{\lambda})|\le\mu^*$ for $d\ge W$. Therefore, $\varepsilon^*=J_0^{-1}(\mu^*)$ should return the minimum value $\varepsilon^*$ such that $|J_0(\varepsilon)|\le\mu^*$ for $\varepsilon\ge\varepsilon^*$, and this is the definition of the inverse used in this paper.

\section{Numerical Results}
In this section, we provide simulation results to understand the outage probability performance of FAS against several important system parameters. To appreciate the capability of FAS, we begin by plotting a typical signal envelope of FAS over time in Figure \ref{fig:afd-t} where we assumed that the FAS is traveling at a speed of $30$km/h operating at frequency of $5$GHz, and the FAS has $100$ ports, with a size of $2\lambda$ (i.e., $12$cm). The signal envelope for a 2-antenna MRC is also provided for comparison. Even though the results only represent one realization of the channel, the channel hardening effect for FAS is obvious, and more pronounced compared to the MRC system. The results also demonstrate that at any given time $\frac{vt}{\lambda}$, the $2\lambda$ space has given a huge range of signal strengths ($50$dB in some cases) from the $100$ spatially correlated ports. This reveals that the limitation due to space at mobile devices may be overstated.

\begin{figure}[]
\begin{center}
\includegraphics[width=15cm]{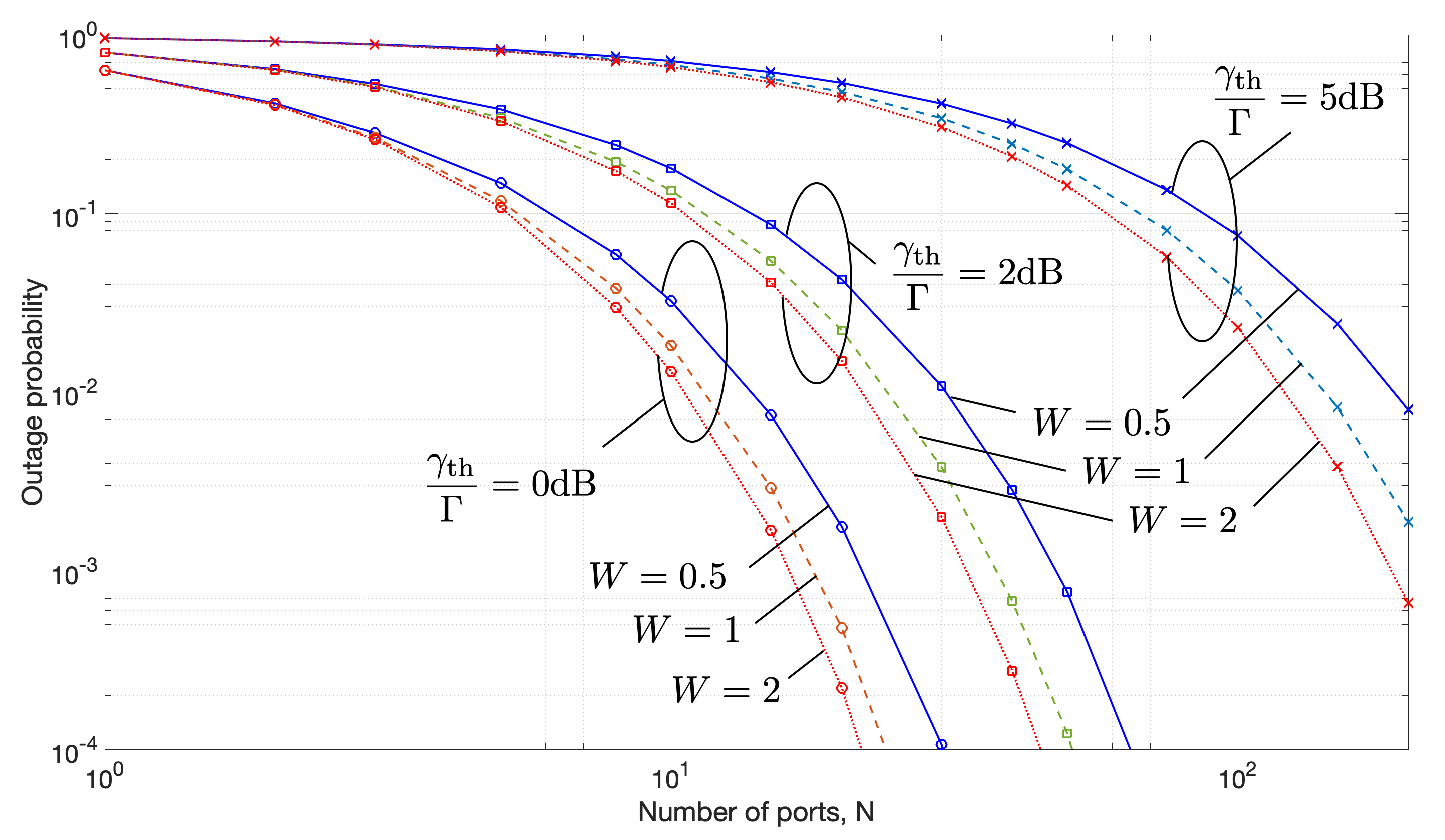}
\caption{Outage probability against $N$ for the FAS.}\label{fig:opVSn}
\end{center}
\end{figure}

In Figure \ref{fig:opVSn}, we investigate how the outage probability performance of FAS scales with the number of ports, $N$, for various sizes, $W$, and SNR targets, $\frac{\gamma_{\rm th}}{\Gamma}$. As expected, as the SNR target becomes more ambitious, the outage probability rises. In addition, if the FAS has more space, i.e., a larger $W$, then the outage probability will be reduced. An important observation in this figure is that the outage probability performance is not limited by the space $W$, and there is no outage floor. As long as $N$ increases, it will continue to drop, which agrees with the statement in Theorem \ref{th:fas-mrc}. There are also a few highlights in the results. We observe that with a tight space of $0.5\lambda$, i.e., $3$cm at $5$GHz, a $10$-port FAS can decrease the outage probability from $0.6$ to $3\times 10^{-2}$ for the case of $\frac{\gamma_{\rm th}}{\Gamma}=10$dB. If we can increase the space to $2\lambda$ and have $20$ ports, the outage probability can even be reduced to $2\times 10^{-4}$, more than $4$ orders of magnitude reduction in the outage probability compared to the single-antenna system without FAS. In short, while space is still an important factor, extraordinary diversity is possible within a tiny space.

\begin{figure}[]
\begin{center}
\includegraphics[width=15cm]{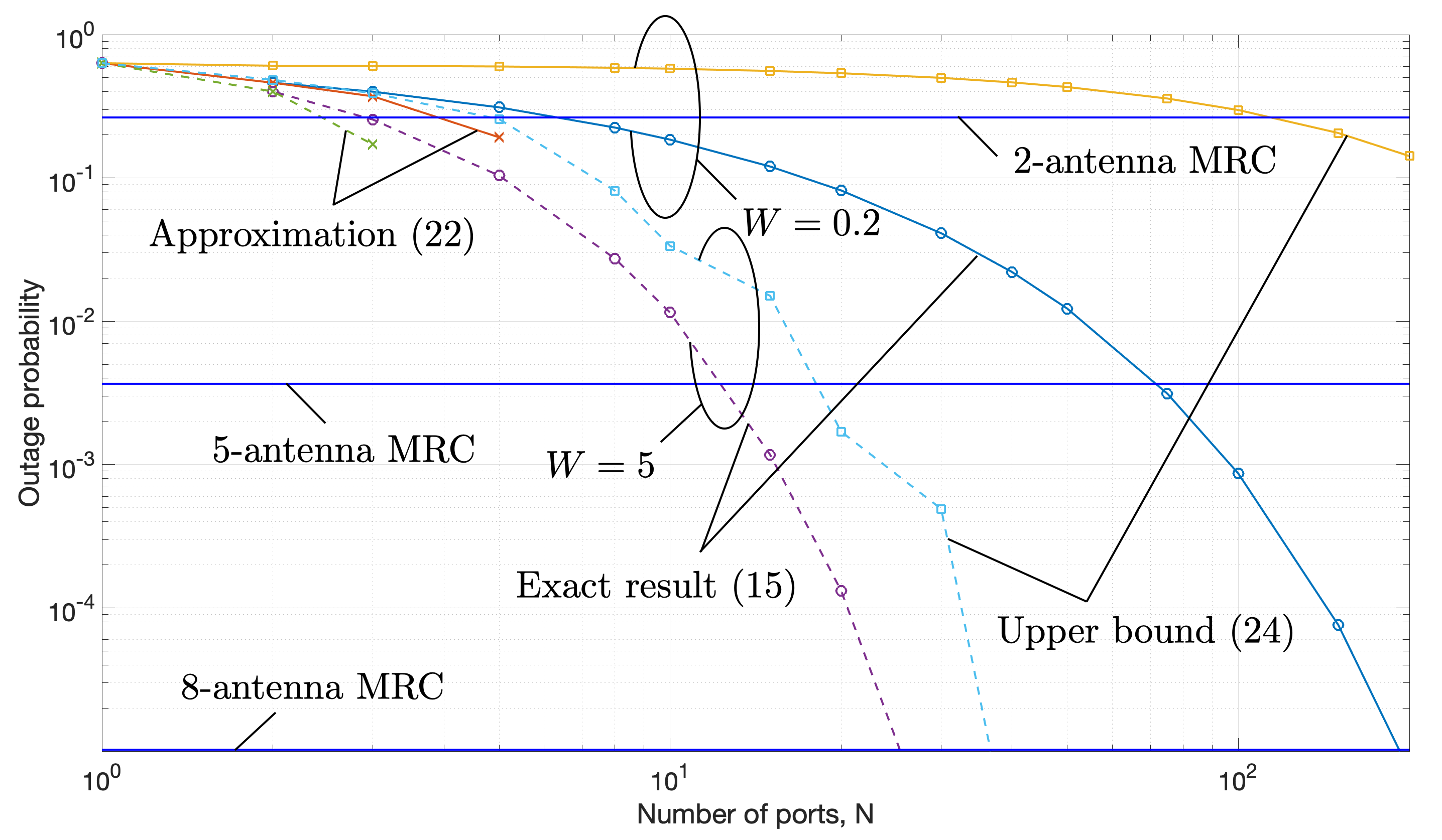}
\caption{Outage probability approximations and bounds for the FAS when $\frac{\gamma_{\rm th}}{\Gamma}=0{\rm dB}$.}\label{fig:opVSn-bounds}
\end{center}
\end{figure}

Figure \ref{fig:opVSn-bounds} assesses the accuracy of the approximation (\ref{eqn:aop}) and the upper bound (\ref{eqn:op-ub}). More importantly, the results in this figure compare FAS with the MRC system with $L$ antennas. In order to show a good range of results, we consider the case of very small $W=0.2$ and the case of large $W=5$.\footnote{Depending on the operating frequency, $W=5$ may still be considered a small space. For example, at $60$GHz, $20\lambda$ space takes up only $10$cm. However, at that frequency, the multipath will be less rich and a more detailed analysis will be required.} First of all, the results illustrate that the approximation (\ref{eqn:aop}) is only accurate when the outage probability is very high. As $N$ increases, (\ref{eqn:aop}) does not scale well and quickly becomes negative and off the chart. This, however, is consistent with the tightness analysis stated in Theorem \ref{th:aop-tight}. On the other hand, the upper bound (\ref{eqn:op-ub}) seems to be good to imitate the falling trend of the outage probability as $N$ increases. This is particularly true when $W=5$, although a large gap is seen when $W$ is too small. While it is fair to say that (\ref{eqn:op-ub}) is not tight, the upper bound (\ref{eqn:op-ub}) plays a key role in the analysis of linking the system parameters with a conservative estimate of the outage probability performance of the FAS.

Now, we compare the performance of FAS with that of MRC. In the figure, we provide the results for $L=2,5,8$. Note that the number of RF chains for MRC needs to match with the number of antennas; yet, in FAS, it always has one RF chain. Remarkably, the results demonstrate that even if we have an extremely small space $W=0.2$, FAS can still outperform a 2-antenna MRC if $N\ge 7$. If $N$ is large enough, say $>70$, then it can even surpass MRC with $5$ antennas. It is worth pointing out that a 5-antenna MRC system with independent fading will need a space of $2\lambda$ which is $10$ times larger than the FAS with $W=0.2$. If $N$ is approaching $200$, the FAS will reach the outage probability of $1\times 10^{-5}$, same as the performance for an 8-antenna MRC, all happening with only a small space of $W=0.2$. If $W$ is larger, the benefit of FAS comes more quickly and FAS can match the performance of an 8-antenna MRC with as small as $N=23$ in the case of $W=5$. Even if $N$ is as small as $12$ which is practically much more possible, a single-antenna FAS can achieve the performance of 5-antenna MRC which is quite remarkable. 

\begin{figure}[]
\begin{center}
\includegraphics[width=15cm]{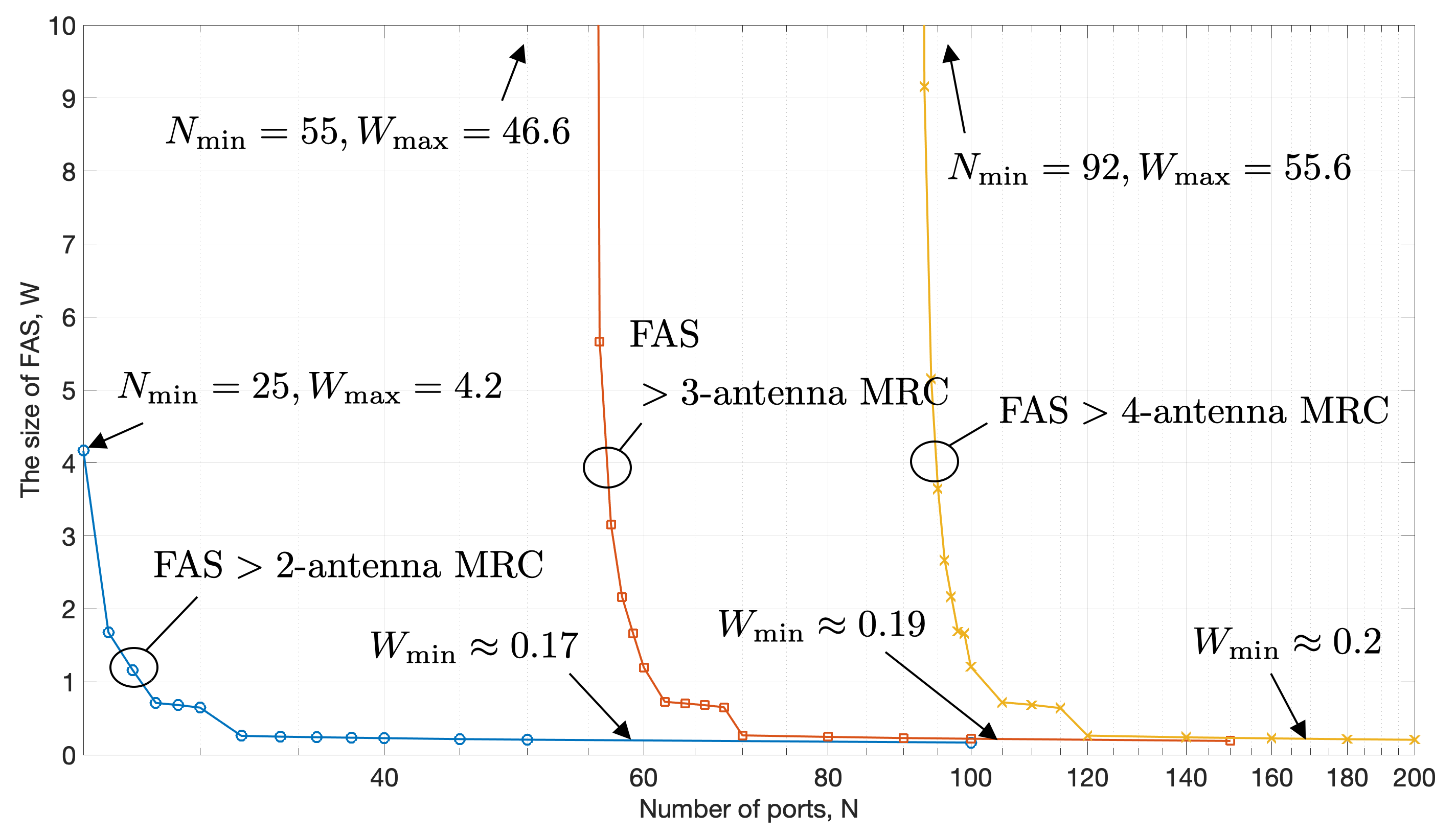}
\caption{The required size of the FAS based on (\ref{eqn:W}) for outperforming the respective MRC system when $\frac{\gamma_{\rm th}}{\Gamma}=0{\rm dB}$. For example, `FAS$>$2-antenna MRC' means that FAS outperforms 2-antenna MRC.}\label{fig:wVSn}
\end{center}
\end{figure}

In Figure \ref{fig:wVSn}, we use (\ref{eqn:W}) to produce the results that can illustrate the relationship between the number of ports, $N$, and the minimum required size, $W$, needed for FAS to outperform MRC based on the upper bound (\ref{eqn:op-ub}). As we can see, there is a minimum number of ports required to make it possible for FAS to surpass MRC and this is true regardless of how large the size of the FAS may be. Despite this, there is a tradeoff between $N$ and $W$. If $N$ is larger, it can make up for the lack of size to achieve the same performance. In particular, results discover that for surpassing 2-antenna MRC, the minimum condition is to have $N=25$ and $W=4.2$. This condition gets much harsher for surpassing 3-antenna MRC, which requires $N=55$ and $W=46.6$. Results also reveal that the required size $W$ decreases sharply though as $N$ increases beyond the critical value. The results in the figure are also useful in estimating the required number of ports for FAS to outperform MRC with a given $W$. For example, for a $1\lambda$-FAS, it will require $N=28$, $N=61$ and $N=102$ ports, respectively, to outperform 2-, 3- and 4-antenna MRC.

We conclude this section by making a few remarks in relation to practical consideration against the above theoretical results. First and foremost, the results should be interpreted with caution, as they are based on the assumption of the model that each port is an ideal point antenna. In practice, each port, effectively an antenna, does take up some physical space. There is also a physical limit in the resolution of ports along the space of a mobile device, which will constrain the feasible values of $N$. It is fair to say that in the extreme values of $N$, the unbelievable performance of FAS may come from the numerical advantage of the model. However, we believe that our results have already illustrated that under practical values of $N$ and $W$, extraordinary performance of FAS is still possible. Moreover, it is foreseeable that metamaterial technologies may make antennas smaller, mimicking the point antenna, which could unlock the physical limitation. More is yet to be explored as fluid antenna technologies mature.

\section*{\sc VI. Conclusion}
The first message this paper wishes to convey is that space being the ultimate limitation for mobile devices is an overstatement and massive diversity is still available in a tiny space. Motivated by the advances in mechanically flexible antennas, this paper studied the concept of FAS where a single {\em fluid antenna} can be switched to the strongest port out of $N$ fixed ports within a given linear space of $W\lambda$. Under an idealized model, we derived the exact and approximate expressions for the outage probability. We also obtained an upper bound of the outage probability from which it has been shown that as $N$ goes to infinity, FAS can achieve any arbitrarily small outage probability regardless of how small $W(>0)$ is. In addition, we used the upper bound to quantify how large $N$ is considered enough for the FAS to outperform an $L$-antenna MRC system. Our numerical results have confirmed the extraordinary capability of FAS given even a small space, and that a FAS with a small space and practically feasible $N$ can outperform MRC. Despite this, there are practical aspects that need further investigation. It is our hope that this paper will spark some interest that leads to a collective effort to unlock the diversity hidden in a small space using FAS.

\section{Appendices}
\subsection{Derivation of $p_{|g_1|,|g_2|,\dots,|g_N|}(r_1,r_2,\dots,r_N)$}
Conditioned on $x_0,y_0$, $|g_2|$ is Rician distributed and therefore, we have
\begin{equation}
p_{|g_2||x_0,y_0}(r_2|x_0,y_0)=\frac{2r_2}{\sigma^2(1-\mu_2^2)}e^{-\frac{r_2^2+\mu_2^2(x_0^2+y_0^2)}{\sigma^2(1-\mu_2^2)}}I_0\left(\frac{\sigma\mu_2\sqrt{x_0^2+y_0^2}r_2}{\frac{\sigma^2}{2}(1-\mu_2^2)}\right),~\mbox{for }r_2\ge 0.
\end{equation}
Given $x_0,y_0$, $|g_2|,|g_3|,\dots,|g_N|$ are all independent and we can obtain
\begin{equation}
p_{|g_2|,\dots,|g_N|||g_1|}(r_2,\dots,r_N|r_1)=\prod_{k=2}^N\frac{2r_k}{\sigma^2(1-\mu_k^2)}e^{-\frac{r_k^2+\mu_k^2r_1^2}{\sigma^2(1-\mu_k^2)}}I_0\left(\frac{\sigma\mu_kr_1r_2}{\frac{\sigma^2}{2}(1-\mu_k^2)}\right),
\end{equation}
where $r_1=\sqrt{x_0^2+y_0^2}$. With the pdf of $|g_1|$ given by (\ref{eqn:|g|}), $p_{|g_2|,\dots,|g_N|||g_1|}(r_2,\dots,r_N|r_1)p_{|g_1|}(r_1)$ gives the desired result (\ref{eqn:jpdf}) which completes the proof.

\subsection{Derivation of $F_{|g_1|,|g_2|,\dots,|g_N|}(r_1,r_2,\dots,r_N)$}
The joint pdf of $|g_1|,|g_2|,\dots,|g_N|$ can be obtained by
\begin{equation}
F_{|g_1|,|g_2|,\dots,|g_N|}(r_1,r_2,\dots,r_N)=\int_0^{r_1}\cdots\int_0^{r_N} p_{|g_1|,|g_2|,\dots,|g_N|}(t_1,t_2,\dots,t_N) dt_1\cdots dt_N.
\end{equation}
Substituting the result in (\ref{eqn:jpdf}) into the above, we get
\begin{equation}
F_{|g_1|,|g_2|,\dots,|g_N|}(r_1,r_2,\dots,r_N)=\int_0^{r_1} \frac{2t_1}{\sigma^2}e^{-\frac{t_1^2}{\sigma^2}} \left[\prod_{k=2}^N\int_0^{r_k}
\frac{2t_k}{\sigma^2(1-\mu_k^2)}e^{-\frac{t_k^2+\mu_k^2t_1^2}{\sigma^2(1-\mu_k^2)}}I_0\left(\frac{2\mu_k t_1t_k}{\sigma^2(1-\mu_k^2)}\right)
dt_k\right] dt_1.
\end{equation}
The integral inside the product operator appears to be an integration over the pdf of a Rician random variable, which therefore gives
\begin{equation}
F_{|g_1|,|g_2|,\dots,|g_N|}(r_1,r_2,\dots,r_N)=\int_0^{r_1} \frac{2t_1}{\sigma^2}e^{-\frac{t_1^2}{\sigma^2}} \prod_{k=2}^N\left[1-Q_1\left(\sqrt{\frac{2\mu_k^2}{\sigma^2(1-\mu_k^2)}}t_1,\sqrt{\frac{2}{\sigma^2(1-\mu_k^2)}}r_k\right)\right] dt_1.
\end{equation}
Now, changing the variable for the integration using $t_1=\frac{r_1^2}{\sigma^2}$ obtains the final result (\ref{eqn:jcdf}).

\subsection{Derivation of the Joint cdf when $N=2$}
To obtain the result, the following lemma is useful.

\begin{lemma}\label{lemma:int2}
It is known that
\begin{equation}\label{eqn:lemma-int2}
\int_0^c e^{-t}Q_1(a\sqrt{t},b)dt=e^{-\frac{b^2}{a^2+2}}Q_1\left(\sqrt{c(a^2+2)},\frac{ab}{\sqrt{a^2+2}}\right)-e^{-c}Q_1(a\sqrt{c},b).
\end{equation}
\end{lemma}

\proof The result can be directly obtained from \cite[(B.19)]{Simon-2002} by simple changes of variables.
\endproof

Substituting $N=2$ in (\ref{eqn:jcdf}), we get 
\begin{align}
F_{|g_1|,|g_2|}(r_1,r_2)&=\int_0^{\frac{r_1^2}{\sigma^2}}e^{-t}\left[1-Q_1\left(\sqrt{\frac{2\mu_2^2}{1-\mu_2^2}}\sqrt{t},\sqrt{\frac{2}{\sigma^2(1-\mu_2^2)}}r_k\right)\right]dt\notag\\
&=1-e^{-\frac{r_1^2}{\sigma^2}}-\int_0^{\frac{r_1^2}{\sigma^2}}e^{-t}Q_1\left(\sqrt{\frac{2\mu_2^2}{1-\mu_2^2}}\sqrt{t},\sqrt{\frac{2}{\sigma^2(1-\mu_2^2)}}r_k\right)dt.
\end{align}
Using (\ref{eqn:lemma-int2}) in Lemma \ref{lemma:int2} on the last term of the above by substituting $a=\sqrt{\frac{2\mu_2^2}{1-\mu_2^2}}$, $b=\sqrt{\frac{2}{\sigma^2(1-\mu_2^2)}}r_k$ and $c=\frac{r_1^2}{\sigma^2}$, we obtain the desired result, and complete the proof.

\subsection{Derivation of the Approximate Outage Probability Expression}
Note that $0\le Q_1(\alpha,\beta)\le 1$ and therefore we can approximate $\prod_k(1-Q_1(\cdot,\cdot))\approx 1-\sum_kQ_1(\cdot,\cdot)$ so that
\begin{align}
p_{\rm out}(\gamma_{\rm th})&\approx\int_0^{\frac{\gamma_{\rm th}}{\Gamma}}e^{-t}\left[1-\sum_{k=2}^NQ_1\left(\sqrt{\frac{2\mu_k^2}{1-\mu_k^2}}\sqrt{t},\sqrt{\frac{2}{1-\mu_k^2}}\sqrt{\frac{\gamma_{\rm th}}{\Gamma}}\right)\right]dt\notag\\
&=1-e^{-\frac{\gamma_{\rm th}}{\Gamma}}-\sum_{k=2}^N\int_0^{\frac{\gamma_{\rm th}}{\Gamma}}e^{-t}Q_1\left(\sqrt{\frac{2\mu_k^2}{1-\mu_k^2}}\sqrt{t},\sqrt{\frac{2}{1-\mu_k^2}}\sqrt{\frac{\gamma_{\rm th}}{\Gamma}}\right)dt.
\end{align}
Using (\ref{eqn:lemma-int2}) in Lemma \ref{lemma:int2} by setting $a=\sqrt{\frac{2\mu_k^2}{1-\mu_k^2}}$, $b=\sqrt{\frac{2}{1-\mu_k^2}}\sqrt{\frac{\gamma_{\rm th}}{\Gamma}}$ and $c=\frac{\gamma_{\rm th}}{\Gamma}$ for the integral inside the summation achieves the approximate expression for the outage probability.

\subsection{Tightness Analysis for (\ref{eqn:aop})}
Before we analyze (\ref{eqn:aop}), the following lemmas are useful.

\begin{lemma}\label{lemma:ubQ}
For $0\le\alpha\le\beta$, $Q_1(\alpha,\beta)$ can be upper-bounded by
\end{lemma}
\begin{equation}\label{eqn:ubQ}
Q_1(\alpha,\beta)<\frac{1}{\sqrt{1+2\alpha\beta}}\left(\frac{\beta}{\beta-\alpha}\right).
\end{equation}

\proof According to \cite{Ferrari-2002}, for $0\le\alpha\le\beta$, we have
\begin{equation}\label{eqn:ubQ-1}
Q_1(\alpha,\beta)\le \frac{I_0(\alpha\beta)}{e^{\alpha\beta}}\left[e^{-\frac{(\beta-\alpha)^2}{2}}+\alpha\sqrt{\frac{\pi}{2}}{\rm erfc}\left(\frac{\beta-\alpha}{\sqrt{2}}\right)\right],
\end{equation}
where ${\rm erfc}(\cdot)$ denotes the complementary error function. Then we apply the upper bounds that ${\rm erfc}(x)\le\frac{e^{-x^2}}{\sqrt{\pi} x}$ \cite{Kschischang-2017} and $I_0(x)<\frac{e^x}{\sqrt{1+2x}}$ \cite[(3.1)]{Yang-2016} into (\ref{eqn:ubQ-1}), giving
\begin{equation}\label{eqn:ubQ-2}
Q_1(\alpha,\beta)<\frac{1}{\sqrt{1+2\alpha\beta}}\left(\frac{\beta}{\beta-\alpha}\right)e^{-\frac{(\beta-\alpha)^2}{2}}<\frac{1}{\sqrt{1+2\alpha\beta}}\left(\frac{\beta}{\beta-\alpha}\right),
\end{equation}
which reaches the desired result in (\ref{eqn:ubQ}).
\endproof

\begin{lemma}\label{lemma:fx}
Given $0\le\mu<1$ and $X_0>\sqrt{\frac{1-\mu^2}{2}}$, define the function $f(x): [0,X_0]\to\mathbb{R}$,
\begin{equation}\label{eqn:fx}
f(x)=\frac{X_0}{(X_0-\mu x)\sqrt{1+\left(\frac{4\mu X_0}{1-\mu^2}\right)x}}>0,~\mbox{for }x\in[0,X_0].
\end{equation}
Then $f(x)$ is a decreasing function as $x$ increases from $0$ and may gradually become an increasing function at some point but once it is increasing, it will continue to increase as $x$ increases. As a result, the maximum of $f(x)$ appears at either $x=0$ or $x=X_0$. Also, $f(0)=1$, $f(X_0)=\frac{1}{(1-\mu)\sqrt{1+\frac{4\mu X_0^2}{1-\mu^2}}}$, and
\begin{equation}
\left.\frac{\partial f(x)}{\partial x}\right|_{x=0}=-\frac{\mu}{1-\mu^2}\left(2X_0-\frac{1-\mu^2}{X_0}\right).
\end{equation}
\end{lemma}

\proof We first obtain the derivative of $f(x)$ as
\begin{equation}\label{eqn:dfx}
\frac{\partial f(x)}{\partial x}=-\frac{\mu X_0\left[2X_0^2-6\mu X_0 x-(1-\mu^2)\right]}{(1-\mu^2)(X_0-\mu x)^2\left(\frac{1-\mu^2+4\mu X_0 x}{1-\mu^2}\right)^\frac{3}{2}}=-\frac{(+{\rm ve})\left[2X_0^2-6\mu X_0 x-(1-\mu^2)\right]}{(+{\rm ve})(+{\rm ve})(+{\rm ve})}.
\end{equation}
Thus, the polarity of $\frac{\partial f(x)}{\partial x}$ is decided by $h(x)=2X_0^2-(1-\mu^2)-6\mu X_0 x$. At $x=0$, $h(0)=2X_0^2-(1-\mu^2)$ and if $X_0>\sqrt{\frac{1-\mu^2}{2}}$, then $h(0)>0$ and $\left.\frac{\partial f(x)}{\partial x}\right|_{x=0}<0$, which shows that $f(x)$ is decreasing from $x=0$. In addition, $h(x)$ is a decreasing function of $x$, meaning that if $x$ is large enough, then $h(x)$ may become negative, making $\frac{\partial f(x)}{\partial x}$ positive, and hence $f(x)$ increasing afterwards. Because of how the value of $f(x)$ changes over $x$, the maximum must appear at one of the endpoints of the interval $[0,X_0]$. Finally, the values of $f(0)$, $f(X_0)$, and $\left.\frac{\partial f(x)}{\partial x}\right|_{x=0}$ can be easily obtained from (\ref{eqn:fx}) and (\ref{eqn:dfx}).
\endproof

Now, we are ready to study the tightness of the approximation (\ref{eqn:aop}). The approximation is based on $\prod_k(1-Q_1(\alpha_k,\beta_k))\approx 1-\sum_kQ_1(\alpha_k,\beta_k)$ which is accurate if $Q_1(\alpha_k,\beta_k)$ is small. In the outage probability computation, $\alpha_k=\sqrt{\frac{2\mu_k^2}{1-\mu_k^2}}\sqrt{t}$ for $0\le t\le\sqrt{\frac{\gamma_{\rm th}}{\Gamma}}$ and $\beta_k=\sqrt{\frac{2}{1-\mu_k^2}}\sqrt{\frac{\gamma_{\rm th}}{\Gamma}}$ and hence, $\alpha_k\le \beta_k$, we can use the upper bound (\ref{eqn:ubQ}) in Lemma \ref{lemma:ubQ} on $Q_1(\alpha_k,\beta_k)$ to yield
\begin{equation}\label{eqn:uQ-3}
Q_1(\alpha_k,\beta_k)<\frac{\sqrt{\frac{\gamma_{\rm th}}{\Gamma}}}{\left(\sqrt{\frac{\gamma_{\rm th}}{\Gamma}}-\mu_k\sqrt{t}\right)\sqrt{1+\frac{4\mu_k\sqrt{\frac{\gamma_{\rm th}}{\Gamma}}}{1-\mu_k^2}\sqrt{t}}},
\end{equation}
where the right-hand-side is recognized as $f(x)$ in (\ref{eqn:fx}) with $x=\sqrt{t}$ and $X_0=\sqrt{\frac{\gamma_{\rm th}}{\Gamma}}$. The upper bound of our interest in (\ref{eqn:uQ-3}) concerns $t\in(0,\frac{\gamma_{\rm th}}{\Gamma})$ (i.e., $x\in(0,X_0)$) in the integration for computing the outage probability (\ref{eqn:op}). Using Lemma \ref{lemma:fx}, the slope of the upper bound of $Q_1(\alpha_k,\beta_k)$ at $t=0$ is given by
\begin{equation}
\left.\frac{\partial f(\sqrt{t})}{\partial \sqrt{t}}\right|_{t=0}=-\frac{\mu_k}{1-\mu_k^2}\left(2\sqrt{\frac{\gamma_{\rm th}}{\Gamma}}-\frac{1-\mu_k^2}{\sqrt{\frac{\gamma_{\rm th}}{\Gamma}}}\right),
\end{equation}
which will be negative and of a very large magnitude if either $\sqrt{\frac{\gamma_{\rm th}}{\Gamma}}$ is large or $\mu_k$ is close to one. If this happens, the upper bound will stay near one only over an insignificant interval of $t$ for the integration. In addition, the upper bound will fall very sharply, and even if it increases again as $t$ increases, the upper bound will be limited by
\begin{equation}
f(X_0)=\frac{1}{(1-\mu_k)\sqrt{1+\left(\frac{4\mu_k}{1-\mu_k^2}\right)\frac{\gamma_{\rm th}}{\Gamma}}},
\end{equation}
which will be small if $\sqrt{\frac{\gamma_{\rm th}}{\Gamma}}$ is large enough. In summary, if $\sqrt{\frac{\gamma_{\rm th}}{\Gamma}}$ and/or $\{\mu_k\}$ are large enough, then the upper bound (and hence $Q_1(\alpha_k,\beta_k)$) will be small over a significant interval of $t$ for the integration (\ref{eqn:op}) and as a consequence, the approximation will be tight, which completes the proof.

\subsection{Derivation of the Upper Bound, $p_{\rm out}^{\rm UB}(\gamma_{\rm th})$}
To derive the upper bound, $p_{\rm out}^{\rm UB}(\gamma_{\rm th})$, it suffices to obtain a lower bound for $\Delta p_{\rm out}(\gamma_{\rm th})$ because according to Theorem \ref{th:op-delta}, $p_{\rm out}^{N}(\gamma_{\rm th})=p_{\rm out}^{N-1}(\gamma_{\rm th})-\Delta p_{\rm out}(\gamma_{\rm th})$ where the superscript $n$ specifies the outage probability for an $n$-port FAS. To proceed, the following lemma is useful.

\begin{lemma}\label{lemma:Q1-lb}
For $0<\alpha<\beta$ and large $\beta$, we have the following lower bound for $Q_1(\alpha,\beta)$:
\begin{equation}\label{eqn:Q1-lb}
Q_1(\alpha,\beta)\gtrsim\varrho\sqrt{\frac{\beta}{\alpha}}e^{-\frac{\kappa}{2}(\beta-\alpha)^2},
\end{equation}
where $\kappa$ is any positive constant greater than one, and $\varrho\triangleq\frac{e^\frac{1}{\pi(\kappa-1)+2}}{2\kappa}\sqrt{\frac{(\kappa-1)(\pi(\kappa-1)+2)}{\pi}}$.
\end{lemma}

\proof For large $\beta$, we have $Q_1(\alpha,\beta)\approx\sqrt{\frac{\beta}{\alpha}}Q(\beta-\alpha)$ \cite[(A.4)]{Simon-2002} in which $Q(x)=\frac{1}{\sqrt{2\pi}}\int_x^\infty e^{-\frac{t^2}{2}} dt$ is the Gaussian $Q$-function. Then applying the lower bound in \cite[Theorem 2.1]{Gross-2012} gives the bound.
\endproof

Using Lemma \ref{lemma:Q1-lb}, for $0<t<\frac{\gamma_{\rm th}}{\Gamma}$ and large $|\mu_N|$, we get
\begin{align}
Q_1\left(\sqrt{\frac{2\mu_N^2}{1-\mu_N^2}}\sqrt{t},\sqrt{\frac{2}{1-\mu_N^2}}\sqrt{\frac{\gamma_{\rm th}}{\Gamma}}\right)&\hspace{+.5mm}\gtrsim\frac{\varrho\left(\frac{\gamma_{\rm th}}{\Gamma}\right)^{0.25}}{t^{0.25}\sqrt{|\mu_N|}}e^{-\frac{\kappa}{1-\mu_N^2}\left(\sqrt{\frac{\gamma_{\rm th}}{\Gamma}}-\mu_N\sqrt{t}\right)^2}\notag\\
&\stackrel{(a)}{>}\frac{\varrho}{\sqrt{|\mu_N|}}e^{-\frac{\kappa}{1-\mu_N^2}\left(\frac{\gamma_{\rm th}}{\Gamma}\right)},\label{eqn:Q1-lb2}
\end{align}
where $(a)$ is obtained by choosing $t=\frac{\gamma_{\rm th}}{\Gamma}$ in the denominator and $t=0$ for the exponential function. As a consequence, we can get a lower bound for $\Delta p_{\rm out}(\gamma_{\rm th})$ by
\begin{align}
\Delta p_{\rm out}(\gamma_{\rm th})&>
\int_0^{\frac{\gamma_{\rm th}}{\Gamma}}e^{-t}\left(\frac{\varrho}{\sqrt{|\mu_N|}}e^{-\frac{\kappa}{1-\mu_N^2}\left(\frac{\gamma_{\rm th}}{\Gamma}\right)}\right)\prod_{k=2}^{N-1}\left[1-Q_1\left(\sqrt{\frac{2\mu_k^2}{1-\mu_k^2}}\sqrt{t},\sqrt{\frac{2}{1-\mu_k^2}}\sqrt{\frac{\gamma_{\rm th}}{\Gamma}}\right)\right]dt\notag\\
&=p_{\rm out}^{N-1}(\gamma_{\rm th})\times\frac{\varrho}{\sqrt{|\mu_N|}}e^{-\frac{\kappa}{1-\mu_N^2}\left(\frac{\gamma_{\rm th}}{\Gamma}\right)},
\end{align}
which then gives
\begin{align}
p_{\rm out}^{N}(\gamma_{\rm th})&<p_{\rm out}^{N-1}(\gamma_{\rm th})\left(1-\frac{\varrho}{\sqrt{|\mu_N|}}e^{-\frac{\kappa}{1-\mu_N^2}\left(\frac{\gamma_{\rm th}}{\Gamma}\right)}\right)\notag\\
&<p_{\rm out}^{N-2}(\gamma_{\rm th})\left(1-\frac{\varrho}{\sqrt{|\mu_{N-1}|}}e^{-\frac{\kappa}{1-\mu_{N-1}^2}\left(\frac{\gamma_{\rm th}}{\Gamma}\right)}\right)\left(1-\frac{\varrho}{\sqrt{|\mu_N|}}e^{-\frac{\kappa}{1-\mu_N^2}\left(\frac{\gamma_{\rm th}}{\Gamma}\right)}\right)\notag\\
&\quad\quad\vdots\notag\\
&<p_{\rm out}^{1}(\gamma_{\rm th})\prod_{k=2}^N\left(1-\frac{\varrho}{\sqrt{|\mu_k|}}e^{-\frac{\kappa}{1-\mu_k^2}\left(\frac{\gamma_{\rm th}}{\Gamma}\right)}\right)=\left(1-e^{-\frac{\gamma_{\rm th}}{\Gamma}}\right)\prod_{k=2}^N\left(1-\frac{\varrho}{\sqrt{|\mu_k|}}e^{-\frac{\kappa}{1-\mu_k^2}\left(\frac{\gamma_{\rm th}}{\Gamma}\right)}\right),
\end{align}
which is the desired result and completes the proof.

\subsection{The Minimum Dimension of Fluid Antenna, $W$}
Corollary \ref{corollary:mu} obtains the minimum required dimension $d^*$ to get the needed autocorrelation $\mu^*$ so that FAS outperforms MRC based on the upper bound of outage probability (\ref{eqn:op-ub}) for a fancy system where all the autocorrelation parameters are the same. Now, imagine, if we have a FAS in which half of the ports, $\frac{N}{2}$, have $d_k<d^*$ but another half have $d_k>d^*$, then this FAS will have better outage probability performance than an $\frac{N}{2}$-port FAS with $\mu_k=\mu^*$. As such, using the performance of the $\frac{N}{2}$-port FAS with $\mu_k=\mu^*~\forall k$ as the worst-case performance for the $N$-port FAS with general $\{\mu_k\}$, we can provide a sufficient condition for the required dimension, $W\lambda$, from the result of Corollary \ref{corollary:mu}.

{\renewcommand{\baselinestretch}{1.1}
\begin{footnotesize}

\end{footnotesize}}

\begin{thebibliography}{1}

\bibitem{Weldon-2014}
H. Viswanathan, and M. Weldon, ``The past, present, and future of mobile communications,'' {\em Bell Labs Tech. J.}, vol. 19, pp. 8--21, Aug. 2014.
\bibitem{Paulraj-1994}
A. J. Paulraj, and T. Kailath, ``Increasing capacity in wireless broadcast systems using distributed transmission/directional reception (DTDR),'' US Patent 5,345,599A, granted 1994.
\bibitem{Foschini-1998}
G. J. Foschini, and M. J. Gans, ``On limits of wireless communications in a fading environment when using multiple antennas,'' {\em Wireless Pers. Commun.}, vol. 6, no. 3, pp. 311--335, Mar. 1998.
\bibitem{Varokh-1998}
V. Tarokh, N. Seshadri, and A. R. Calderbank, ``Space-time codes for high data rate wireless communication: Performance criterion and code construction,'' {\em IEEE Trans. Inform. Theory}, vol. 44, no. 2, pp. 744-765, Mar. 1998.
\bibitem{Alamouti-1998}
S. M. Alamouti, ``A simple transmit diversity technique for wireless communications,'' {\em IEEE J. Select. Areas Commun.}, vol. 16, no. 8, pp. 1451--1458, Oct. 1998.
\bibitem{Tse-2003}
L. Zheng, and D. N. C. Tse, ``Diversity and multiplexing: A fundamental tradeoff in multiple-antenna channels,'' {\em IEEE Trans. Inform. Theory}, vol. 49, no. 5, pp. 1073--1096, May 2003.
\bibitem{Goldsmith-2003}
S. Vishwanath, N. Jindal, and A. Goldsmith, ``Duality, achievable rates, and sum-rate capacity of Gaussian MIMO broadcast channels,'' {\em IEEE Trans. Inform. Theory}, vol. 49, no. 10, pp. 2658--2668, Oct. 2003.
\bibitem{Spencer-2004}
Q. H. Spencer, A. L. Swindlehurst, and M. Haardt, ``Zero-forcing methods for downlink spatial multiplexing in multiuser MIMO channels,'' {\em IEEE Trans. Signal Proc.}, vol. 52, no. 2, pp. 461--471, Feb. 2004.
\bibitem{Marzetta-2013}
H. Q. Ngo, E. G. Larsson, and T. L. Marzetta, ``Energy and spectral efficiency of very large multiuser MIMO systems,'' {\em IEEE Trans. Commun.}, vol. 61, no. 4, pp. 1436--1449, Apr. 2013.
\bibitem{Marzetta-2014}
E. G. Larsson, O. Edfors, F. Tufvesson, and T. L. Marzetta, ``Massive MIMO for next generation wireless systems,'' {\em IEEE Commun. Mag.}, vol. 52, no. 2, pp. 186--195, Feb. 2014.
\bibitem{Stuber-2002}
G. L. St$\ddot{\rm u}$ber, {\em Principles of Mobile Communication}, Second Edition, Kluwer Academic Publishers, 2002.

\bibitem{Randy-2013}
R. L. Haupt, and M. Lanagan, ``Reconfigurable antennas,'' {\em IEEE Antennas and Propag. Mag.}, vol. 55, no. 1, pp. 49--61, Feb. 2013.

\bibitem{Hayes-2012}
G. J. Hayes, J.-H. So, A. Qusba, M. D. Dickey, and G. Lazzi, ``Flexible liquid metal alloy (EGaIn) microstrip patch antenna,'' {\em IEEE Trans. Antennas Propag.}, vol. 60, no. 5, pp. 2151--2156, May 2012.
\bibitem{Ohta-2013}
A. M. Morishita, C. K. Y. Kitamura, A. T. Ohta, and W. A. Shiroma, ``A liquid-metal monopole array with tunable frequency, gain, and beam steering,'' {\em IEEE Antennas Wireless Propag. Lett.}, vol. 12, pp. 1388--1391, 2013.
\bibitem{Saghati-2014}
A. P. Saghati, J. Batra, J. Kameoka, and K. Entesari, ``A microfluidically-tuned dual-band slot antenna,'' in Proc. {\em IEEE Antennas Propag. Soc. Int. Symp. (APSURSI)}, pp. 1244--1245, 6-11 Jul. 2014, Memphis, TN, USA.
\bibitem{Dey-2016}
A. Dey, R. Guldiken, and G. Mumcu, ``Microfluidically reconfigured wideband frequency-tunable liquid-metal monopole antenna,'' {\em IEEE
Trans. Antennas Propag.}, vol. 64, no. 6, pp. 2572--2576, Jun. 2016.
\bibitem{Tong-2017}
C. Borda-Fortuny, K.-F. Tong, A. Al-Armaghany, and K. K. Wong, ``A low-cost fluid switch for frequency-reconfigurable Vivaldi antenna,'' {\em IEEE Antennas Wireless Propag. Lett.}, vol. 16, pp. 3151--3154, 2017.
\bibitem{Tong-2018}
C. Borda-Fortuny, K. F. Tong, and K. Chetty, ``Low-cost mechanism to reconfigure the operating frequency band of a Vivaldi antenna for cognitive radio and spectrum monitoring applications,'' {\em IET Microwaves, Antennas \& Propag.}, vol. 12, no. 5, pp. 779--782, 2018. 
\bibitem{Singh-2019}
A. Singh, I. Goode, and C. E. Saavedra, ``A multistate frequency reconfigurable monopole antenna using fluidic channels,'' {\em IEEE Antennas Wireless Propag. Lett.}, vol. 18, no. 5, pp. 856--860, May 2019.
\bibitem{Tong-2019}
C. Borda-Fortuny, L. Cai, K. F. Tong, and K. K. Wong, ``Low-cost 3D-printed coupling-fed frequency agile fluidic monopole antenna system,'' {\em IEEE Access}, pp. 95058--95064, Jul. 2019.
\bibitem{Soh-2020}
K. N. Paracha, A. D. Butt, A. S. Alghamdi, S. A. Babale, and P. J. Soh, ``Liquid metal antennas: Materials, fabrication and applications,'' {\em Sensors 2020}, 20, 177.
\bibitem{Murch-2014}
S. Song, and R. D. Murch, ``An efficient approach for optimizing frequency reconfigurable pixel antennas using genetic algorithms,'' {\em IEEE Trans. Antennas Propag.}, vol. 62, no. 2, pp. 609--620, Feb. 2014.


\bibitem{Neil-2019}
N. Convery, and N. Gadegaard, ``30 years of microfluidics,'' {\em Micro and Nano Engineering}, vol. 2, pp. 76--91, Mar. 2019.
\bibitem{Ugweje-1997}
O. C. Ugweje, and V. Aalo, ``Performance of selection diversity system in correlated Nakagami fading,'' in Proc. {\em IEEE Veh. Technol. Conf.  (VTC)}, pp. 1488--1492, 4-7 May 1997,  Phoenix, AZ, USA.
\bibitem{Slim-2003}
L. Yang, and M.-S. Alouini, ``An exact analysis of the impact of fading correlation on the average level crossing rate and average outage duration of selection combining,'' in Proc. {\em  IEEE Veh. Technol. Conf. (VTC-Spring)}, pp. 241--245, 22-25 Apr. 2003, Jeju, South Korea.
\bibitem{Simon-2002}
M. K. Simon, {\em Probability Distributions Involving Gaussian Random Variables: A Handbook for Engineers and Scientists}, Springer, Boston, MA, 2002.

\bibitem{Ferrari-2002}
G. E. Corazza, and G. Ferrari, ``New bounds for the Marcum Q-function,'' {\em IEEE Trans. Inform. Theory}, vol. 48, pp. 3003--3008, Nov. 2002.
\bibitem{Kschischang-2017}
F. R. Kschischang, {\em The complementary error function}, Available [Online]: https://www.comm.utoronto.ca/frank/notes/erfc.pdf.
\bibitem{Yang-2016}
Z.-H. Yang, and Y.-M. Chu, ``On approximating the modified Bessel function of the first kind and Toader-Qi mean,'' {\em J. Inequalities and Appl.}, vol. 40, 2016.
\bibitem{Gross-2012}
F. D. C\^{o}t\'{e}, I. N. Psaromiligkos, and W. J. Gross, ``A Chernoff-type lower bound for the Gaussian Q-function,'' [Online]. Available: https://arxiv.org/abs/1202.6483.


\end{thebibliography}
\end{document}